\documentclass[prl,longbibliography,twocolumn,superscriptaddress,amsmath,amssymb,verbatim]{revtex4-1}

\usepackage{graphicx}
\usepackage{lmodern}
\usepackage{epstopdf}
\usepackage[colorlinks=true,citecolor=blue,linkcolor=blue,urlcolor=blue,bookmarks=true]{hyperref}

\renewcommand{\eqref}[1]{Eq.~(\ref{#1})}

\begin{document}

\preprint{APS/123-QED}

\title{Symmetry Reduction in the Quantum Kagome Antiferromagnet Herbertsmithite}

\author{A. Zorko}
\email{andrej.zorko@ijs.si}
\affiliation{Jo\v{z}ef Stefan Institute, Jamova c.~39, SI-1000 Ljubljana, Slovenia}
\author{M. Herak} 
\email{mirta@ifs.hr}
\affiliation{Institute of Physics, Bijeni\v{c}ka c.~46, HR-10000 Zagreb, Croatia}
\author{M. Gomil\v sek}
\affiliation{Jo\v{z}ef Stefan Institute, Jamova c.~39, SI-1000 Ljubljana, Slovenia}
\author{J. van Tol}
\affiliation{National High Magnetic Field Laboratory, Florida State University, Tallahassee, Florida 32310, USA}
\author{M. Vel\'azquez}
\affiliation{CNRS, Universit\'e de Bordeaux, ICMCB, UPR 9048, 87 Avenue du Dr.~A.~Schweitzer, 33608 Pessac Cedex, France}
\author{P. Khuntia}
\affiliation{Laboratoire de Physique des Solides, CNRS, Univ.~Paris-Sud, Universit\'e Paris-Saclay, 91405 Orsay Cedex, France}
\author{F. Bert}
\affiliation{Laboratoire de Physique des Solides, CNRS, Univ.~Paris-Sud, Universit\'e Paris-Saclay, 91405 Orsay Cedex, France}
\author{P.~Mendels}
\affiliation{Laboratoire de Physique des Solides, CNRS, Univ.~Paris-Sud, Universit\'e Paris-Saclay, 91405 Orsay Cedex, France}

\date{\today}

\begin{abstract}
Employing complementary torque magnetometry and electron spin resonance on single crystals of herbertsmithite, the closest realization to date of a quantum kagome antiferromagnet featuring a spin-liquid ground state, we provide novel insight into different contributions to its magnetism.
At low temperatures, two distinct types of defects with different magnetic couplings to the kagome spins are found. 
Surprisingly, their magnetic response contradicts the threefold symmetry of the ideal kagome lattice, suggesting the presence of a global structural distortion that may be related to the establishment of the spin-liquid ground state.     
\end{abstract}

\maketitle

Spin-induced breaking of crystal symmetry is a widespread phenomenon in one-dimensional antiferromagnets \cite{pytte1974peierls,bray1975observation, hase1993observation}.
There, spin degrees of freedom can conspire to form singlets, i.e., valence bonds, leading to lattice dimerization known as the spin-Peierls transition.
In higher dimensions, a similar lattice instability induced by magnetoelastic coupling can appear in geometrically frustrated magnets to relieve frustration by lifting the degeneracy of magnetically ordered ground states  \cite{lacroix2011introduction, lee2000local, becca2002peierls, chung2005statics, giot2007magnetoelastic, zorko2014frustration, zorko2015magnetic}.
A more intriguing option though is a spin-Peierls-like transition due to spin paring in disordered but correlated spin states of frustrated lattices.
The latter phenomenon was indeed theoretically predicted \cite{yamashita2000spin,tchernyshyov2002spin,hermanns2015spin} and also found experimentally in a triangular-lattice compound \cite{tamura2006frustration} and a Shastry-Sutherland lattice representative \cite{vecchini2009structural}, both featuring spin-singlet ground states.
The two-dimensional quantum kagome antiferromagnet (QKA), a paradigm of correlated disordered spin states, seems to be resistant against valence-bond ordering, as a spin-liquid (SL) ground state preserving the lattice symmetry is predicted \cite{yan_spin-liquid_2011, depenbrock_nature_2012, jiang_identifying_2012}.
However, this state is only slightly energetically favorable as compared to a valence-bond crystal (VBC) \cite{singh2007ground}, which breaks the translational symmetry and could lead to a lattice distortion if assisted by the magnetoelastic coupling. 
Since defects, inherently present in kagome materials, may be able to pin a VBC \cite{poilblanc2010impurity}, it is interesting to pose the question of how much such perturbations to the QKA can modify its ground state. 

From early investigations \cite{bert2007low, olariu200817, de2008magnetic}, randomness in the form of Zn-Cu intersite disorder is an issue well known for the so far most intensively studied QKA representative herbertsmithite, ZnCu$_3$(OH)$_6$Cl$_2$.
The amount of disorder is sizable, as 5\%-8\% of the Cu$^{2+}$ ions end up at the Zn intersite even for the best polycrystalline samples and single crystals \cite{bert2007low, olariu200817, de2008magnetic, freedman2010site}.
Additionally, the presence of Zn$^{2+}$ vacancies at the intralayer kagome Cu site was suggested from early $^{17}$O nuclear magnetic resonance (NMR) experiments \cite{olariu200817}, but was later disputed based on x-ray anomalous scattering \cite{freedman2010site} and NMR measurements on single crystals \cite{fu2015evidence}.
The defects behave like weakly coupled spin-1/2 entities \cite{mendels_quantum_2010} and represent the dominant contribution to thermodynamic quantities at low energies and low temperatures ($T$'s) \cite{de2009scale, han2012fractionalized, nilsen2013low, han2015correlated}. 
A recent inelastic neutron scattering study has helped to disentangle the kagome and defect excitation spectra and has revealed that the defects are correlated \cite{han2015correlated}.
The important question of how much they affect the
SL ground state of the kagome spins, however, remains unsettled \cite{han2015correlated, kawamura2014quantum, kermarrec2011spin}.
\begin{figure*}[t]
\includegraphics[trim = 0mm 0mm 0mm 0mm, clip, width=1\linewidth]{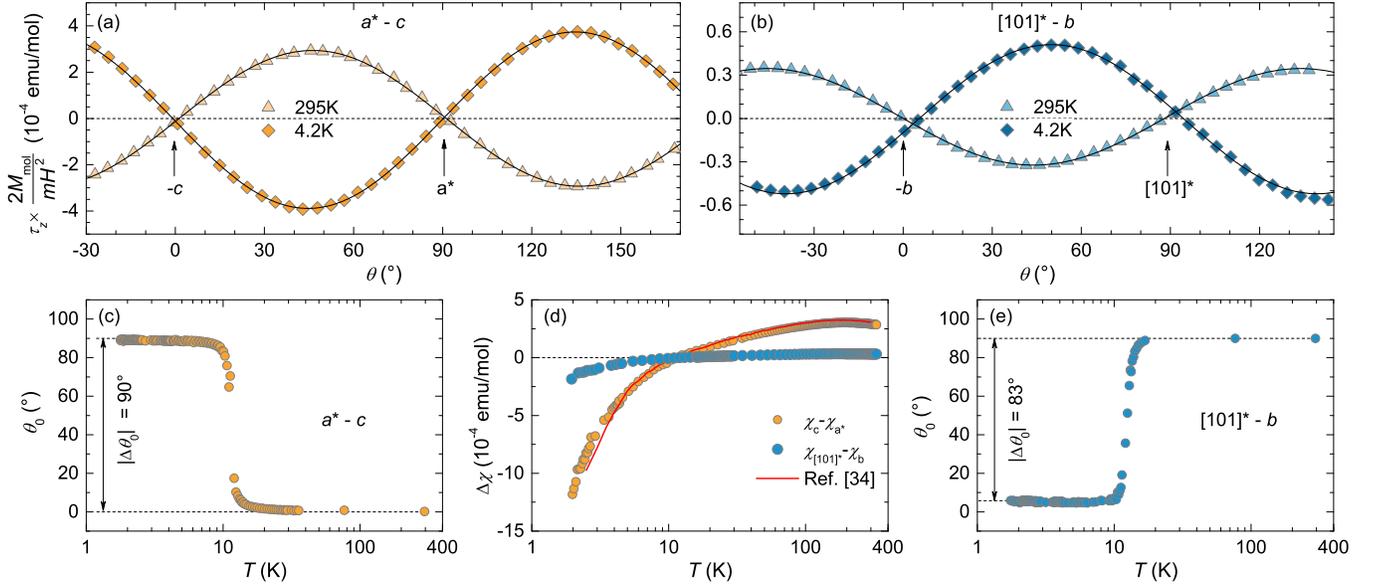}
\caption{
(a), (b) The angular dependence of the magnetic torque $\tau_z$ in two different crystallographic planes. 
The solid lines represent fits with Eq.~(\ref{eq1}).
The $T$ dependence of (c), (e) the corresponding magnetic-torque phase $\theta_0$ and (d) the corresponding magnetic anisotropy for both crystallographic planes. 
The solid line in (d) represents the magnetization data published in Ref.~\onlinecite{han2012refining}.
The dashed lines are a guide to the eye.
In (e) they highlight the magnetic-torque phase change of $|\Delta \theta_0| \neq 90^\circ$, which reveals broken symmetry with $\chi_{a^*}\neq \chi_b$ (see text for details).
}
\label{fig1}
\end{figure*}

Here we provide a novel insight into the problem of the interplay of defects and the kagome physics  in herbertsmithite by combining bulk magnetic torque and local-probe electron spin resonance (ESR) measurements on single crystals.
The torque magnetometry surpasses conventional magnetization measurements when determining the orientation of the principal axes of the magnetic susceptibility tensor, while high resolution and sensitivity of ESR allow individual inspection of different contributions to magnetism if these differ in $g$ factors or linewidths \cite{zorko2013dzyaloshinsky}.
Indeed, in herbertsmithite, we clearly demonstrate with ESR the presence of two distinct types of defects that prevail at low $T$'s.
Unexpectedly, the corresponding $g$-factor anisotropy as well as the $T$-dependent orientation of the magnetic axes provide complementary evidence that both defect contributions break the threefold symmetry of the kagome lattice, which reveals a global structural distortion. 

In our investigation we used several single crystals of masses $m$=5--17~mg with irregular  shapes, similar to the one reported in Ref.~\onlinecite{han2012refining}.
The crystals were grown following the published procedure \cite{han2011synthesis}.
The high quality of the samples was verified by x-ray diffraction (XRD) and SQUID magnetometry.
The direction of the crystallographic axes was determined by XRD using a Laue camera in backscattering geometry.

The magnetic torque $\tau_z$ was measured along the $z$ laboratory axis on a home-built apparatus in a fixed magnetic field of $\mu_0H=0.2$~T ($\mu_0$ is the vacuum permeability) rotating in the $xy$ laboratory plane \cite{sup}.
This complements previous high-field torque measurements that were performed for fixed directions of $H$ as a function of the field strength \cite{asaba2014high}.
In our experiment, the field was rotated in two different planes, the first one containing the $a^*$ and $c$ crystallographic axes [Fig.~\ref{fig1}(a)], and the second one comprising the $[101]^*$ and $b$ axes [Fig.~\ref{fig1}(b)].
Here, $c$ denotes the axis perpendicular to the kagome planes.
All measured angular dependences of $\tau_z$ are typical of systems with a magnetic response linear in $H$, where \cite{sup}
\begin{equation}\label{eq1}
	\tau_z = \dfrac{m}{2 M_{\rm mol}} H^2  \left( \chi_{x'} - \chi_{y'}\right)\sin (2\theta - 2\theta_0).
\end{equation}
Here, $M_{\rm mol}$ is the molar mass of the sample, $\chi_{x'}$ and $\chi_{y'}$ are the maximal and minimal susceptibility values within the $xy$ plane, $\theta$ is the angle between the $H$ and $x$ axis, while the torque phase $\theta_0$ represents the angle between the $x'$ and $x$ axes. 
From the torque amplitude one obtains the susceptibility anisotropy $\Delta \chi_{xy} = \chi_{x'} - \chi_{y'}$ in the measured $xy$ plane while $\theta_0$ tracks the rotation of the magnetic easy axis within the $xy$ plane. 

From measurements in the $a^*$-$c$ plane [Fig.~\ref{fig1}(a)], we find that at high $T$'s, i.e., for the intrinsic kagome spins that dominate the magnetic response of herbertsmithite above $\sim$100~K \cite{mendels_quantum_2010}, the magnetic easy axis is oriented along $c$ ($\chi_{c}>\chi_{a*}$).
Around the crossover temperature of $T_0\approx 12$~K, $\theta_0$ changes by 90$^\circ$ [Fig.~\ref{fig1}(c)], revealing that $\chi_{c}<\chi_{a*}$ for the defect spins, which are dominant at low $T$'s.
Thus the bulk easy axis $c$ is replaced by the bulk easy plane $ab$ below $T_0$, in agreement with previous bulk susceptibility measurements [Ref.~\onlinecite{han2012refining}; see Fig.~\ref{fig1}(d)].
At first glance, similar information is obtained from measurements within the $[101]^*$-$b$ plane [Fig.~\ref{fig1}(b)]. 
The sign of $\chi_{[101]^*}-\chi_{b}$ changes from positive to negative at $T_0$ on decreasing $T$ [Fig.~\ref{fig1}(d)], as expected, because the $[101]^*$ direction contains a small part of the $c$ component.
However, unexpectedly, the torque curves for 4.2 and 295~K cross zero at notably different $\theta_0$ [Fig.~\ref{fig1}(b)], revealing a total change of the torque phase $|\Delta\theta_0|=83^\circ$ around $T_0$ [Fig.~\ref{fig1}(e)].
This change is incompatible with threefold crystal symmetry yielding $\chi_{a^*}=\chi_{b}$, which can only support the torque-phase change of $90^\circ$ \cite{sup}. 
The measurements thus irrefutably disclose a deviation from the uniaxial symmetry.
Note that, for randomly and independently distributed defects, an average uniaxial global symmetry should be preserved in bulk magnetic measurements, even if defects possess lower local symmetry. 
A torque phase change that differs from 90$^\circ$ is thus a fingerprint of a reduced crystal symmetry.

\begin{figure}[t]
\includegraphics[trim = 0mm 2mm 0mm 0mm, clip, width=1\linewidth]{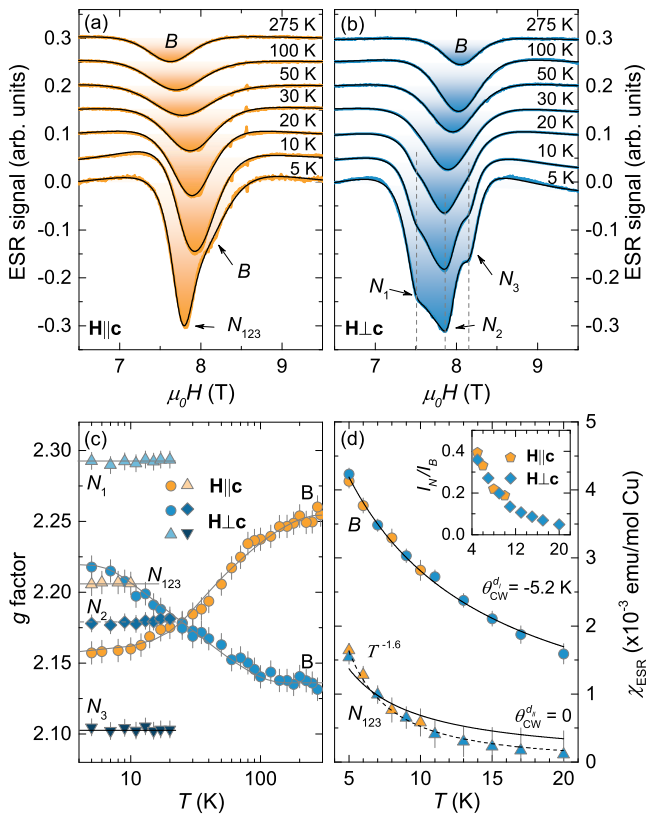}
\caption{
The ESR spectra for (a) {\bf H}$||${\bf c} and (b) {\bf H}$\perp${\bf c}, which
are displaced vertically for clarity.
The fits (solid lines) disclose a single broad ($B$) ESR component at high $T$'s, while  an additional narrow component ($N_{123}$) is found for {\bf H}$||${\bf c} and three additional components ($N_1$, $N_2$ and $N_3$) for {\bf H}$\perp${\bf c} at low $T$'s. The positions of the latter are indicated by dashed lines.
(c) The $T$ dependence of the $g$ factors of individual components.
The lines are a guide to the eye.
(d) The ESR susceptibility of the $B$ and $N$ components and their measured ratio (inset).
The solid lines show fits with the Curie-Weiss model, the dashed line corresponds to a power-law dependence. 
}
\label{fig2}
\end{figure}

In order to further inspect this surprising bulk symmetry reduction, we performed complementary ESR measurements at the NHMFL, Tallahassee, Florida on a custom made ESR spectrometer operating at 240~GHz.
A single broad ESR line (labeled $B$ in Fig.~\ref{fig2}) was observed at high $T$'s for {\bf H}$||${\bf c} and {\bf H}$\perp${\bf c}, with the linewidth and $g$ factors \cite{sup} in agreement with previous data from polycrystalline samples  \cite{zorko2008dzyaloshinsky}.
The ESR intensity $I(T)$, which is proportional to the magnetic susceptibility $\chi_{\rm ESR}$ \cite{kubo1954general}, was previously found to scale with the bulk susceptibility rather than with the kagome susceptibility in powder samples \cite{zorko2008dzyaloshinsky}.
This remains the case for single crystals [Fig.~\ref{fig2}(d)].
The corresponding crossover of the ESR spectra from high $T$'s (kagome spins) to low $T$'s (defects, labeled $d_I$) is clearly observed by the variation of the ESR linewidths \cite{sup,zorko2008dzyaloshinsky} and $g$ factors [Fig.~\ref{fig2}(c)] between 100 and 10~K.
The change of the $g$ factors corroborates the crossover from the high-$T$ $c$ easy axis to the low-$T$ $ab$ easy-plane-like behavior also found by the torque magnetometry. 
We note that the existence of a single broad ESR line, instead of separated lines from the kagome spins and the defect spins, is due to exchange narrowing \cite{bencini1990electron} and reveals that the coupling $J^{d_I}$ between the two spin species is larger than $J_H=\Delta g\mu_{\rm B}\mu_0H/k_{\rm B}\approx 2$~K, where $\Delta g\sim 0.3$ is the $g$-factor anisotropy typical for Cu$^{2+}$ \cite{abragam19702electron}, $\mu_{\rm B}$ the Bohr magneton, and $k_{\rm B}$ the Boltzman constant.

In contrast to powder ESR spectra \cite{zorko2008dzyaloshinsky}, additional, narrower components appear at low $T$'s in single crystals -- 
a single narrow component ($N_{123}$) below $\sim$10~K for {\bf H}$||${\bf c} [Fig.~\ref{fig2}(a)] and three similarly intense narrow components ($N_1$, $N_2$ and $N_3$) below $\sim$20~K for {\bf H}$\perp${\bf c} [Fig.~\ref{fig2}(b)]. 
Since the relative intensity of the $N$ components with respect to the broad $B$ component is the same for the two field orientations, $I_{N_{123}}/I_{B}=(I_{N_1}+I_{N_2}+I_{N_3})/I_{B}$ [inset in Fig.~\ref{fig2}(d)], a single additional defect site in the crystal structure of herbertsmithite is required.
The ESR intensity of these new defects (labeled ${d_{II}}$) amounts to $\sim$40\% of the $B$ component at 5~K, where the latter is mainly assigned to ${d_I}$ defects. Therefore, we can write $\chi_{\rm ESR}^{B}=\chi^{\rm k}+\chi_{\rm ESR}^{d_I}$ for the B component, where $\chi^{\rm k}$ denotes the intrinsic kagome susceptibility,
 while for the N component $\chi_{\rm ESR}^{N}=\chi_{\rm ESR}^{d_{II}}$.
After separating the two defect contributions \cite{sup},
we find [Fig.~\ref{fig2}(d)] that $\chi_{\rm ESR}^{d_I}\propto1/(T-\theta_{\rm CW}^{d_I})$, with the Weiss temperature $\theta_{\rm CW}^{d_I}=-5.2$~K, in agreement with the above conclusion $J^{d_I}\gtrsim J_H$.
In contrast, $\chi_{\rm ESR}^{d_{II}}$ increases much faster with decreasing $T$, possibly even faster than the Curie law ($\theta_{\rm CW}^{d_{II}}=0$), as a phenomenological $T^{-p}$ ($p=1.6$) dependence works even better [Fig.~\ref{fig2}(d)].   
This, together with the much smaller linewidth of the $N$ components \cite{sup} and the fact that they are not exchange narrowed for {\bf H}$\perp${\bf c}, reveals that the exchange coupling $J^{d_{II}}$ associated with the ${d_{II}}$ defects is small, i.e., $|J^{d_{II}}|\ll J_H$.
\begin{figure}[b]
\includegraphics[trim = 0mm 0mm 0mm 0mm, clip, width=0.95\linewidth]{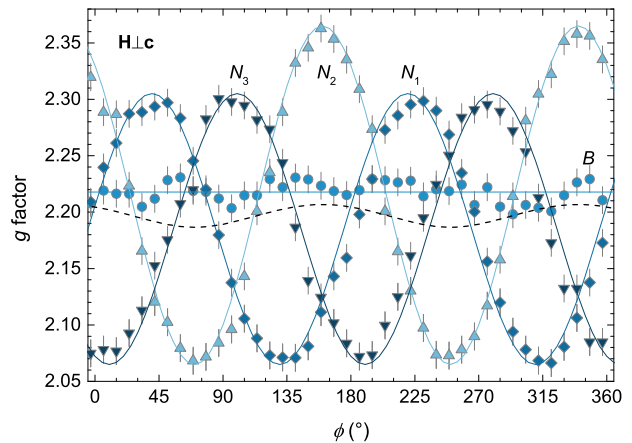}
\caption{
The angular dependence of the $g$ factors for {\bf H}$\perp${\bf c} at 5~K.
The $B$ component is angle independent (horizontal line), while the $N$ components behave according to the model $g_j(\phi)=\left[g_{j,\rm min}^2\cos^2(\phi-\phi_j)+g_{j,\rm max}^2\sin^2(\phi-\phi_j) \right]^{1/2}$ (solid lines).
The dashed line corresponds to the average $\sum_j g_j(\phi)/3$ and demonstrates the breaking of uniaxial symmetry.
The {\bf H}$\perp${\bf c} measurements in Fig.~\ref{fig2}(b) were taken at $\phi=30^\circ$.
}
\label{fig3}
\end{figure}    

Additional insight into the two defect contributions at low $T$'s is provided by the angular dependence of the $g$ factors within the kagome planes at 5~K (Fig.~\ref{fig3}).
The $B$ component is angle independent within the experimental uncertainty, as would be expected for the threefold symmetry of the kagome lattice.
The three $N$ components, corresponding to the $d_{II}$ defects, each show the expected $180^\circ$ periodicity and a relative shift of $60^\circ$.
However, as one of the three components exhibits a notably larger $g$-factor anisotropy than the other two, the threefold symmetry of the crystal structure is evidently broken. 
This is the second unambiguous fingerprint of symmetry reduction in herbertsmithite.
For the $N$ defect component we find an average $g$-factor anisotropy of $\Delta g_{ab}^{d_{II}}=0.02$ within the kagome planes (see the dashed line in Fig.~\ref{fig3}), while for the $B$ defect component we can estimate its upper bound, $\Delta g_{ab}^{d_{I}}\lesssim 0.01$.

Further assessments of $\Delta g_{ab}^{d_{I}}$ can be done based on simulations of the magnetic-torque phase $\theta_0$ \cite{sup}.
At high $T$'s, the torque is determined by the susceptibility of the kagome spins, while below $T_0$ it is mainly given by defects.
Since $\chi_{\rm ESR}^{d_I}\gg\chi_{\rm ESR}^{d_{II}}$ at $T_0\approx 12$~K [inset in Fig.~\ref{fig2}(d)], where $\theta_0$ abruptly changes [Figs.~\ref{fig1}(c),~(e)], only the dominant ${d_I}$ defect contribution is considered.
If $g_{a}^{d_{I}}=g_{b}^{d_{I}}$, which would be in accordance with the threefold kagome-lattice symmetry, only a change of $|\Delta\theta_0|=90^\circ$ could occur in any crystallographic plane \cite{sup} at the temperature where $\Delta\chi_{xy}^{\rm k}=\Delta\chi_{xy}^{d_{I}}$. 
Allowing $g_{a}^{d_{I}}\neq g_{b}^{d_{I}}$ leads to $|\Delta\theta_0|\neq 90^\circ$ for the $[101]^*$-$b$ plane while leaving $\Delta\theta_0=90^\circ$ for the $a^*c$ plane, because $c$ remains a magnetic eigenaxis \cite{sup}.
This is in perfect agreement with the experiment [Fig.~\ref{fig1}(c), (e)].
Within this model, the experimentally observed phase change of $|\Delta\theta_0|=83^\circ$ is reproduced for $\Delta g_{ab}^{d_{I}}\sim 0.003(1)$ \cite{sup}.
Such a small $g$-factor anisotropy, nevertheless revealing a symmetry reduction also for the broad ESR component, can obviously not be resolved by ESR.
The magnetic-torque measurements thus nicely complement ESR measurements by exposing the symmetry reduction also for the other type of detects. 
We note that 
both the unexpected torque-phase change [Fig.~\ref{fig1}(e)] and the $g$-factor irregularity (Fig.~\ref{fig3}) were observed in all investigated crystals.
Thus, the symmetry reduction reflected in the magnetism of defects is obviously an intrinsic feature of herbertsmithite.

It seems reasonable to attribute the $d_I$ defects with a sizable Weiss temperature $\theta_{\rm CW}^{d_I}=-5.2$~K to Cu$^{2+}$ spins at the Zn intersites, which are expected to be coupled to the kagome spins.
Namely, the exchange couplings between the kagome layers that amount to several kelvins run through the intersites \cite{jeschke2013first}. 
Moreover, pairs of such defects are correlated across the kagome layers at low temperatures \cite{han2015correlated}.
On the other hand, the $d_{II}$ defects are rather exceptional, as they appear quasifree ($\theta_{\rm CW}^{d_{II}}\sim 0$).
Furthermore, they differ substantially in $g$ factors from the $d_I$ defects [Fig.~\ref{fig2}(c)], implying a different local environment.
The Zn$^{2+}$ vacancies at the kagome sites, if present, could potentially be responsible for such defects by inducing local magnetization patterns in the surrounding spin liquid \cite{poilblanc2010impurity, dommange2003static, rousochatzakis2009dzyaloshinskii}.
If such vacancies are absent \cite{freedman2010site} one could reconcile the existence of the $d_{II}$ defects by a local effect the interplane Cu$^{2+}$ spins might have on the kagome planes by perturbing the intrinsic spin liquid.
Such a local perturbation can pin a spinon kagome-lattice excitation \cite{kolezhuk2006theory}, as recently found in the related kagome compound Zn-brochantite \cite{Gomilsek2016muSR, Gomilsek2016instabilities}. 

Trivially, Cu$^{2+}$ at the Zn intersites could profoundly affect the intrinsic kagome physics by inducing random bonds on the kagome lattice \cite{kawamura2014quantum}.
However, the required amount of disorder is unlikely realized in herbertsmithite and there is no apparent reason why the {\it global} threefold crystal symmetry should be broken.
The same should apply to {\it local} Jahn-Teller distortions of the perfect octahedral environment at the Zn intersites induced by the Cu$^{2+}$ spins \cite{norman2016herbertsmithite}. 
Then, the important question arises of whether the observed global symmetry reduction is in any way related to the establishment of the SL ground state, which could provide a global driving force.
Namely, the symmetry reduction in herbertsmithite may arise from local distortions if these are effectively coupled through the correlated electronic state of the system.
In this case, the lattice distortion should appear around the temperature where strong spin correlations pertinent to the SL state develop, i.e. around 50~K, where the kagome susceptibility exhibits its maximal value \cite{olariu200817,imai2011local,fu2015evidence}.
Even though our investigation could not directly determine the possible onset temperature of this distortion -- this would require unfeasible magnetic-torque measurement within the kagome plane \cite{sup} -- 
further arguments that favor this scenario can be found in the enhanced $^{35}$Cl NMR relaxation rate at 50~K \cite{imai2008cu, fu2015evidence} and in a pronounced change of the quadrupolar frequency between 50 and 100~K at the $^{17}$O site next to defects \cite{fu2015evidence}.

The discovery of the symmetry reduction in herbertsmithite is interesting in the context of a striped spin-liquid-crystal state, which was recently proposed as an instability of the Dirac SL on the kagome lattice that breaks the uniaxial lattice symmetry, and also reduces the gauge symmetry from $U(1)$ to $Z_2$ \cite{clark2013striped}.
The only synchrotron XRD report at low $T$'s published to date failed to detect any obvious structural phase transition in herbertsmithite \cite{han2011synthesis}. However, the rather small values of $\Delta g_{ab}$ suggest that deformation of the hexagonal structure may be very small.
On the other hand, a recent infrared reflectivity investigation reported an anomalous low-$T$ broadening of a low-frequency phonon mode \cite{sushkov2016infrared}.
Although this was interpreted as possibly signaling $p6$ chirality symmetry breaking  \cite{capponi2013p} in the SL ground state, it could also be related to a structural deformation.   

In conclusion, our study has revealed that two types of intrinsic magnetic defects exist in herbertsmithite at low $T$'s, which differ significantly in their interactions with the surrounding kagome spins. 
The strongly exchange-coupled defects are identified by the broad ESR component with a Curie-Weiss $T$ dependence of the susceptibility, while the second defect type, corresponding to the narrow ESR components, is characterized by much weaker coupling and a steeper increase of their susceptibility on decreasing $T$, possibly beyond the Curie model.
Explaining the nature of the unexpectedly narrow ESR response of the latter defects and their apparent isolation from the kagome spins should be a key avenue in future studies of the interplay between the defect and the intrinsic kagome physics in herbertsmithite.
Moreover, both types of defects have provided evidence of a global symmetry reduction of the kagome lattice in herbertsmithite at low $T$'s.
This changes the perspective on herbertsmithite and may have implications for the selection of its ground state. 
In order to better understand this intriguing discovery, refined {\it ab initio} calculations of relaxed structures in the presence and absence of the intersite disorder could potentially be highly informative, since the energy difference between perfect and defect structures is minimal \cite{guterding2016prospect}.

%
\begin{acknowledgments}
We are grateful to K.~Salamon from the Institute of Physics in Zagreb for performing Laue diffraction. We acknowledge the financial support of the Slovenian Research Agency through projects BI-HR/14-15-003 and BI-FR/15-16-PROTEUS-004 and program No.~P1-0125. M.~H.~acknowledges the support of the Croatian Ministry of Science, Education and Sports and full financial support of the work performed in Zagreb by the Croatian Science Foundation under the project UIP-2014-09-9775.
P.~K.~acknowledges support from the European Commission through Marie Curie International Incoming Fellowship (PIIF-GA-2013-627322).
This work was also supported by the French Agence Nationale de la
Recherche under ``SPINLIQ" Grant No. ANR-12-BS04-0021 and by Universit\'e Paris-Sud Grant MRM PMP.
NHMFL is supported by the NSF through the cooperative agreement DMR-1157490, the State of Florida and the Department of Energy.
\end{acknowledgments}
%
%

%
\newpage
\begin{widetext}
\vspace{19cm}
\begin{center}
{\large {\bf Supplemental information:\\Symmetry Reduction in the Quantum Kagome Antiferromagnet Herbertsmithite}}\\
\vspace{0.5cm}
A. Zorko,$^{1, *}$ M. Herak,$^{2, \dagger}$ M. Gomil\v sek,$^{1}$ J. van Tol,$^{3}$ M. Vel\'azquez,$^{4}$ P. Khuntia,$^{5}$ F. Bert,$^{5}$ and P. Mendels$^{5}$
\vspace{0.3cm}

{\it \small
$^1$Jo\v{z}ef Stefan Institute, Jamova c.~39, SI-1000 Ljubljana, Slovenia\\
\vspace{0cm}
$^2$Institute of Physics, Bijeni\v{c}ka c.~46, HR-10000 Zagreb, Croatia\\
\vspace{0cm}
$^3$National High Magnetic Field Laboratory, Florida State University, Tallahassee, Florida 32310, USA\\
\vspace{0cm}
$^4$CNRS, Universit\'e de Bordeaux, ICMCB, UPR 9048,\\
 87 Avenue du Dr.~A.~Schweitzer, 33608 Pessac Cedex, France\\
\vspace{0cm}
$^5$Laboratoire de Physique des Solides, CNRS, Univ.~Paris-Sud,\\
 Universit\'e Paris-Saclay, 91405 Orsay Cedex, France\\
}

\end{center}
\end{widetext}
\section{Torque Magnetometry and Simulations}\label{appA}
In magnetic-torque measurements the samples were fixed between two quartz plates of the sample holder, one fixed and perpendicular to the plane of rotation of the magnetic field and the other connected via a quartz spring. Limited by the largest facets of the samples and their irregular shapes, the samples were mounted to lie with the crystallographic (101) plane on the fixed quartz plate and could be rotated around the $[101]^*$ axis. In this way the torque could be measured within the $a^*$-$c$ and $[101]^*$-$b$ crystallographic planes.
The planes were determined at room temperature by reorienting the sample in small steps until the torque amplitude was maximized/minimized for the $a^*$-$c$ plane/$[101]^*$-$b$ plane.
This gave an initial orientational accuracy of about $\pm 5^{\circ}$.
At the same time the torque curve had to have phase shifts $\theta_0=0$ and $\theta_0=90^\circ$  for the $a^*$-$c$ and the $[101]^*$-$b$ plane, respectively.
With this condition an accuracy better than $\pm 2^{\circ}$ was reached.
The most likely misalignments that could occur in our experiment were those related to rotations around the [101]* axis, because the (101) plane corresponding to the largest facet was well defined.
After each temperature-dependent measurement $\theta_0$ and the torque amplitude were remeasured again at room temperature to ensure that the orientation of the sample had not changed.
Absolute orientation of the magnetic field with respect to the sample holder with maximal uncertainty of $0.3^\circ$ was determined separately before the experiment.

Torque magnetometry is a method, which can be used to determine the magnetic anisotropy and the orientation of bulk magnetic axes with respect to crystallographic axes.
Contrary to bulk susceptibility measurements where the magnetic field is applied along a specific crystallographic direction, torque measurements, where the field is rotated within a chosen plane, can directly detect rotations of magnetic axes with temperature.
The magnetic torque $\boldsymbol{\tau}$ acting on a sample of volume $V$ and magnetization $\mathbf{M}$ placed in the magnetic field $\mathbf{H}$ is given by
$\boldsymbol{\tau} = V \mathbf{M} \times \mathbf{H}$.
If the induced magnetization is linearly proportional to the applied field, $\mathbf{M} = {\underline{\boldsymbol\chi}} \cdot \mathbf{H}$, the magnetic torque along the $z$ direction is given by
\begin{equation}\label{eq:torquez}
	\tau_z = \dfrac{m}{2 M_{\rm mol}} H^2 \left[ \left( \chi_{xx} - \chi_{yy}\right)\sin 2\theta - 2\chi_{xy} \cos 2\theta \right],
\end{equation}
when the magnetic field is rotated withing the $xy$ plane and $\theta$ denotes the angle between $H$ and a chosen $x$ laboratory axis. Here, $m$ is the mass and $M_{\rm mol}$ the molar mass of the sample, while 
\begin{equation}\label{eq:tensorchi}
	\underline{\bm{\chi}} = 
	\begin{bmatrix}
	\chi_{xx} & \chi_{xy} & \chi_{xz}\\
	\chi_{xy} & \chi_{yy} & \chi_{yz} \\
	\chi_{xz} & \chi_{yz} & \chi_{zz}
	\end{bmatrix}
\end{equation}
is the general susceptibility tensor in the laboratory frame of reference. With the use of relations
\begin{subequations}
\begin{align}
\label{eq:theta0}
\tan(2\theta_0)&=\dfrac{2\chi_{xy}}{\chi_{xx}-\chi_{yy}},\\
\label{eq:tensorcomp}
\chi_{x'} - \chi_{y'} &= \dfrac{\chi_{xx}-\chi_{yy}}{\cos(2\theta_0)}=\dfrac{2\chi_{xy}}{\sin(2\theta_0)},
\end{align}
\end{subequations}
Eq.~(\ref{eq:torquez}) is transformed to
\begin{equation}\label{eq:torquemj}
	\tau_z = \dfrac{m}{2 M_{\rm mol}} H^2  \left( \chi_{x'} - \chi_{y'}\right)\sin (2\theta - 2\theta_0),
\end{equation}
where $\chi_{x'}$ and $\chi_{y'}$ are the maximal and minimal susceptibility values within the $xy$ plane, while $\theta_0$ is the angle between the $x'$ easy axis and the $x$ axis.
If we denote the susceptibility tensor in its eigenframe by 
\begin{equation}\label{eq:tensorchidia}
	\underline{\bm{\chi}} = 
	\begin{bmatrix}
	\chi_{X} & 0 & 0\\
	0 & \chi_{Y} & 0 \\
	0 & 0 & \chi_{Z}
	\end{bmatrix}
\end{equation}
Eq.~(\ref{eq:torquez}) will transform to
\begin{eqnarray}\label{eq:taucosines2}\nonumber
	\tau_{z} &\propto&  [(\alpha_{xX}^2 - \alpha_{yX}^2)(\chi_{X}-\chi_{Z}) +\\\nonumber
	&+& (\alpha_{xY}^2 - \alpha_{yY}^2) (\chi_{Y}-\chi_{Z}) ]\sin(2\theta) - \\  \nonumber
	&-& 2 [\alpha_{xX} \alpha_{yX}(\chi_{X}-\chi_{Z}) +\\
	&+& \alpha_{xY} \alpha_{yY} (\chi_{Y} - \chi_{Z})] \cos(2\theta),
\end{eqnarray}
where $\alpha_{ij}$ are direction cosines of the eigenvectors in the laboratory frame. Taking into account Eq.~(\ref{eq:theta0}) we derive
\begin{multline}\label{eq:theta0cosines}
	 \tan(2\theta_0) = \\
	 2\dfrac{\alpha_{xX} \alpha_{yX}(\chi_{X}-\chi_{Z}) + \alpha_{xY} \alpha_{yY}(\chi_{Y}-\chi_{Z})}{(\alpha_{xX}^2-\alpha_{yX}^2)(\chi_{X}-\chi_{Z})+(\alpha_{xY}^2-\alpha_{yY}^2)(\chi_{Y}-\chi_{Z})}.
\end{multline}
 
Due to the threefold rotational symmetry with its axis parallel to the $c$ axis that should be obeyed by both the intrinsic-kagome susceptibility and the defect-spin susceptibility tensors in herbertsmithite, and hence also by the total susceptibility tensor, $\chi_{X}=\chi_{Y}=\chi_a$, $\chi_{Z}=\chi_c$. This simplifies Eq.~(\ref{eq:theta0cosines}) to
\begin{equation}\label{eq:theta0herbz}
	\tan(2\theta_0) = \dfrac{2\alpha_{xc}\alpha_{yc}}{\alpha_{xc}^2-\alpha_{yc}^2}.
\end{equation}
Eq.~(\ref{eq:theta0herbz}) demonstrates that $\tan(2\theta_0)$ should be a constant number during $T$-dependent measurements in any crystallographic plane. Therefore,
the torque phase $\theta_0$ cannot be $T$-dependent in uniaxial systems, aside from discrete jumps by $\pm 90^\circ$ at a temperature $T_0$, where the easy axis $x'$ and the hard axis $y'$ within the $xy$ plane are interchanged [Eq.~(\ref{eq:torquemj})].
Any misalignment of the sample thus plays no role. 
The phase shift of $90^\circ$ is indeed observed for the $a^*$-$c$ plane [Fig.~1(c) in the main text], while for the $[101]^*$-$b$ plane the torque-phase $\theta_0$ changes by significantly less than $90^\circ$ [Fig.~1(e) in the main text]. 

\begin{figure}[t]
\includegraphics[trim = 0mm 0mm 0mm 0mm, clip, width=1\linewidth]{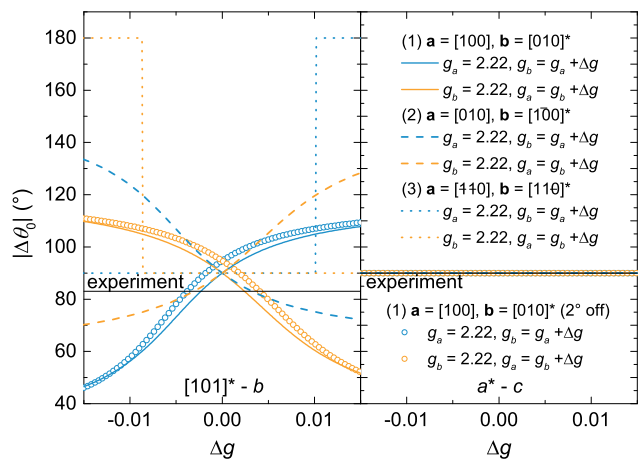}
\caption{
The simulated magnetic-torque-phase change $\Delta\theta_0$ between high and low temperatures in herbertsmithite for the two experimentally investigated crystal planes. Three different orientation of the magnetic axis within the $ab$ plane and two possible directions of the easy axis for each case are considered (thick lines). 
Symbols show simmulations for $2^\circ$ sample misalignment.
The thin (black) solid lines corresponds to the experiment.  
All simulations collapse on the experimental line for the $a^*$-$c$ plane (right panel). 
}
\label{figS1}
\end{figure}

In order to explain the experimentally observed changes of $\theta_0$, we performed magnetic-torque simulations.
At high temperatures, where the kagome spins dominate, we presume uniaxial symmetry , $\chi_{a}^{\rm k}=\chi_{b}^{\rm k}\neq\chi_{c}^{\rm k}$, while at low temperatures, where defects become dominant,  a reduction of this symmetry is allowed, $\chi^{d_{I}}_a\neq\chi^{d_{I}}_b \neq\chi^{d_{I}}_c$.
We assume $\chi^{\rm k}_{c,a^*}\propto (g^{\rm k}_{c,a^*})^2$, with the $g$-factor eigenvalues $g_{c}^{\rm k}=2.26$ and $g_{a}^{\rm k}=2.13$, as inferred from ESR at high $T$'s [Fig.~2(c) in the main text].
Taking into account the bulk susceptibility at room temperature \cite{han2012refiningSI}, we calculate $\Delta\chi_{c,a^*}=2.6\times 10^{-4}$~emu/mol, which is in excellent agreement with the torque measurements [Fig.~1(d) in the main text]. 
Similarly, for defects we have $\chi^{d_{I}}_{c,a^*}\propto (g^{d_{I}}_{c,a^*})^2$, where the two $g$-factor eigenvalues $g_{c}^{d_{I}}=2.16$ and $g_{a}^{d_{I}}=2.22$ are taken from the broad ESR component at low $T$'s.
We considered three possibilities for the eigenaxes direction within the kagome planes at low $T$'s; (1) ${\bf a}=[100]$, ${\bf b}=[010]^*$, (2) ${\bf a}=[010]$, ${\bf b}=[\bar{1}00]^*$, and (3) ${\bf a}=[\bar{1}\bar{1}0]$, ${\bf b}=[1\bar{1}0]^*$, each with the two possible directions of the easy axis within the plane, $\chi_a>\chi_b$ and $\chi_a<\chi_b$ (Fig.~\ref{figS1}).
The orientation of the magnetic axes and the direction of the easy axis within the kagome plane significantly affect the magnetic-torque curves.
The experimentally observed torque-phase change between high and low temperatures of $|\Delta\theta_0|=83^\circ$ in the $[101]^*$-$b$ plane is compatible with two considered orientations of the magnetic axes and requires $|\Delta g_{ab}^{d_{I}}|\sim 0.003$ [Fig.~\ref{figS1}(a)].
For the $a^*$-$c$ plane, on the other hand, the torque-phase change between high and low temperatures remains $\Delta\theta_0=90^\circ$ even for $\chi^{d_{I}}_a\neq\chi^{d_{I}}_b$, in agreement with experiment.

The maximal possible misalignment of $\pm 2^\circ$ introduces an uncertainty in the evaluated $|\Delta g_{ab}^{d_{I}}|$. 
The corresponding curve in Fig.~\ref{figS1}(a) for one of the possible magnetic-axes orientations yields $|\Delta g_{ab}^{d_{I}}|\sim 0.003(1)$. 
A similar result is obtained for the other possible orientation.
At the same time, such misalignment does not affect the torque shift for the measurements within the $a^*$-$c$ plane [Fig.~\ref{figS1}(b)], in line with Eq.~(\ref{eq:theta0herbz}).

The difference between $\chi^{d_{I}}_a$ and $\chi^{d_{I}}_b$ of course requires a finite magnetic torque even within the $ab$ crystallographic (kagome) plane.
Measurement within this plane should, therefore, directly reveal the temperature at which the symmetry reduction of the crystal structure occurs in herbertsmithite.
Unfortunately, due to irregular shapes of the crystals, measurements within the $ab$ plane were not feasible.
\begin{figure}[t]
\includegraphics[trim = 0mm 0mm 0mm 0mm, clip, width=1\linewidth]{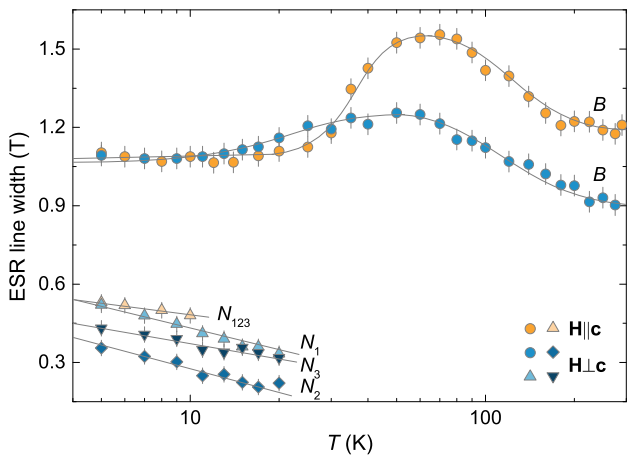}
\caption{
The $T$ dependence of the ESR linewidths in herbertsmithite for the broad (B) component and the narrow (N) components. The solid lines are a guide to the eye.
}
\label{figS2}
\end{figure}  

\section{Electron Spin Resonance}\label{appB}
The ESR measurements were performed on a transmission-type heterodyne ESR spectrometer equipped with a goniometer at a fixed microwave frequency of $\nu=240$~GHz by applying an ac magnetic-field modulation of 2~mT at a frequency of 42~kHz in addition to a sweeping static magnetic field $H$.
The samples were placed into 2-mm diameter quartz tubes filled with vacuum grease, which allowed the sample to reorient in the magnetic field.
The tubes were put into a 12-T magnetic field at room temperature, with the tubes either parallel or perpendicular to the field.
The samples were found to orient in the magnetic field with the $c$ crystallographic axis (corresponding to the largest $g$-factor eigenvalue) along the direction of the field.
At temperatures where the measurements were performed the grease solidifies and ensures the same orientation of the sample is kept at all temperatures.
By minimizing/maximizing the resonance field at 200~K, the crystallographic planes either including or excluding the $c$ crystallographic axis could be selected.
Any misalignment of the samples was estimated to be $\lesssim 2^\circ$. 

In Fig.~2 of the main text the dispersion part of the spectra is shown. 
At high $T$'s a single derivative Lorentzian function was fit to the spectra to extract the linewidth and the $g$ factor, $g=h\nu/\mu_{\rm B}\mu_0 H$, where $h$ denotes the Planck constant, $\mu_{\rm B}$ the Bohr magneton and $\mu_0$ the vacuum permeability.
The observed linewidths [Fig.~\ref{figS2}], which are due to exchange anisotropy and scale with $J$ \cite{bencini1990electronSI}, and $g$ factors [Fig.~2(c) in the main text] are the same in single crystals as previously derived from powder-sample simulations \cite{zorko2008}.
Even though three magnetically nonequivalent kagome sites exist for a general orientation of the magnetic field, a single broad ESR line arises due to exchange narrowing \cite{bencini1990electronSI} when $J\gtrsim J_H=\Delta g\mu_{\rm B}\mu_0H/k_{\rm B}\approx 2$~K, where $\Delta g\sim 0.3$ is $g$-factor anisotropy typical for Cu$^{2+}$ \cite{abragam19702electronSI} and $k_{\rm B}$ is the Boltzman constant.

In contrast to powder spectra \cite{zorko2008}, at low $T$'s, additional narrower components appear in the ESR spectra of single crystals, either one for {\bf H}$||${\bf c} or three for {\bf H}$\perp${\bf c} (Fig.~2 in the main text).
The relative intensity of the three lines for the latter case,  $(I_{N_1}+I_{N_2}+I_{N_3})/I_{B}$, is the same as of the single narrow line for the former case, $I_{N_{123}}/I_{B}=I_{N}/I_{B}$ [inset in Fig.~2(d) in the main text].
The $T$ dependence of the corresponding linewidths is shown in Fig.~\ref{figS2}.    

The ESR intensity is proportional to the magnetic susceptibility, $I(T)=c(T) \chi_{\rm ESR}(T)$ \cite{zorko2008}, where the proportionality constant $c(T)$ is, in principle, $T$-dependent.
However, from the known ratio $I_{N}/I_{B}(T)$ the magnetic susceptibilities of the individual ESR component can be extracted.
We postulate (see main text) $\chi_{\rm ESR}^{B}=\chi^{\rm k}+\chi_{\rm ESR}^{d_I}$ and $\chi_{\rm ESR}^{N}=\chi_{\rm ESR}^{d_{II}}$, where $\chi^{\rm k}$ corresponds to the susceptibility of kagome spins, while $\chi_{\rm ESR}^{d_I}$ and $\chi_{\rm ESR}^{d_{II}}$ are the susceptibilities of defects, yielding the broad and narrow ESR components at low $T$'s, respectively.
Additionally, the three individual contributions sum up to the total bulk susceptibility of the system, $\chi^{\rm bulk}=\chi^{\rm k}+\chi_{\rm ESR}^{d_I}+\chi_{\rm ESR}^{d_{II}}$.
Then,
\begin{subequations}
\begin{align}
\label{eq:ESRint1}
\chi_{\rm ESR}^{d_I}&=\frac{\chi^{\rm bulk}}{\frac{I_{N}}{I_{B}}+1}-\chi^{\rm k},\\
\label{eq:ESRint2}
\chi_{\rm ESR}^{d_{II}}&=\chi^{\rm bulk}\frac{\frac{I_{N}}{I_{B}}}{\frac{I_{N}}{I_{B}}+1}.
\end{align}
\end{subequations}
The bulk susceptibility of herbertsmithite is well known \cite{mendels_quantum_2010SI,han2012refiningSI}, while we extracted $\chi^{\rm k}$ from NMR data in Ref.~\onlinecite{fu2015evidenceSI}, by assuming no major anisotropy of the latter.
The resulting magnetic susceptibilities $\chi_{\rm ESR}^{d_I}$ and $\chi_{\rm ESR}^{d_{II}}$ of the two defect types are displayed in Fig.~2(d) of the main text.

\end{document}